\documentclass[8pt, twocolumn]{article}

\usepackage{amsmath}
\usepackage{amssymb}
\usepackage{graphicx}
\usepackage{color}
\usepackage{multicol}
\usepackage{lipsum}
\usepackage[sort&compress]{natbib}

\usepackage{bm}
\usepackage{authblk}

\usepackage{geometry}
\usepackage{mdframed}
\usepackage{tcolorbox}

\usepackage[english]{babel}

\newcommand{\cost}{\mathcal{C}}
\newcommand{\risk}{\mathcal{R}}

\newcommand{\dd}{\mathrm{d}}
\newcommand{\avg}{\mathbb{E}}

\newcommand{\abs}{\textrm{abs}}
\newcommand{\rel}{\textrm{rel}}

\newcommand{\bQ}{{\bm Q}}

\newcommand{\bq}{{\bm q}}
\newcommand{\bx}{{\bm x}}

\newcommand{\bp}{{\bm p}}
\newcommand{\bpi}{{\bm \pi}}
\newcommand{\bpsi}{{\bm \psi}}
\newcommand{\bsgm}{{\bm \sigma}}
\newcommand{\liq}{g}

\newcommand{\mG}{{\bm {\mathsf{G}}}}
\newcommand{\mSgm}{{\bm {\mathsf{\Sigma}}}}
\newcommand{\mRho}{{\bm {\mathsf{\rho}}}}
\newcommand{\mO}{{\bm {\mathsf{O}}}}
\newcommand{\mLam}{{\bm {\mathsf{\Lambda}}}}
\newcommand{\mLiq}{{\bm {\mathsf{g}}}}

\newcommand{\mR}{{\bm {\mathsf{r}}}}
\newcommand{\mC}{{\bm {\mathsf{c}}}}

\newcommand{\dphi}{\dot\phi}

\newcounter{boxonecounter}
\newenvironment{boxoneequation}{%
\addtocounter{equation}{-1}
\refstepcounter{boxonecounter}

\begin{equation}}
{\end{equation}}

\newenvironment{boxoneeqnarray}[1]{%
\addtocounter{equation}{-#1}
\refstepcounter{boxonecounter}

\begin{eqnarray}}
{\end{eqnarray}}

\newcounter{boxtwocounter}
\newenvironment{boxtwoequation}{%
\addtocounter{equation}{-1}
\refstepcounter{boxtwocounter}

\begin{equation}}
{\end{equation}}

\newenvironment{boxtwoeqnarray}[1]{%
\addtocounter{equation}{-#1}
\refstepcounter{boxtwocounter}

\begin{eqnarray}}
{\end{eqnarray}}

\begin{document}

\title{Trading Lightly: \\Cross-Impact and Optimal Portfolio Execution}
\author[1]{I~Mastromatteo}
\author[1,2]{M~Benzaquen}
\author[1]{Z~Eisler}
\author[1]{J-P~Bouchaud}
\affil[1]{Capital Fund Management, 23 rue de l'Universit\'e, 75007 Paris}
\affil[2]{Ladhyx, UMR CNRS 7646, \'Ecole Polytechnique, 91128 Palaiseau
  Cedex, France}
\renewcommand\Authands{ and }

\twocolumn[
  \begin{@twocolumnfalse}
    \maketitle
    \begin{abstract}
      We model the impact costs of a strategy that trades a basket of correlated instruments,
      by extending to the multivariate case the linear propagator model previously used for
      single instruments. Our specification allows us to calibrate a cost model that is free of arbitrage and price
      manipulation. We illustrate our results using a pool
      of US stocks and show that neglecting cross-impact effects leads to an incorrect estimation of the liquidity and
      suboptimal execution strategies. We show in particular the importance of synchronizing the execution of correlated 
      contracts.
    \end{abstract}
  \end{@twocolumnfalse}
]

Executing trading decisions in real markets is a difficult business. Moving around substantial amounts is often 
foiled by the lack of liquidity. When orders exceed the small volumes typically
available at the best bid or ask in lit order books, the incurred slippage costs rapidly become detrimental to the trading strategy.
Accurately predicting these trading costs is quite a non-trivial exercise, and one must often resort to statistical models.
Most of the complexity of such models arises from {\it price impact}, i.e. the fact that trades tend to move market
prices in their own direction. This effect has recently triggered a large amount of
theoretical activity, due to its direct relation to supply and demand, and also because of the abundantly available data on
financial transactions. As alluded to above, this interest is not purely academic, since reliable estimates of trading costs
are crucial in order to judge whether one should enter into a position and -- if the cost model is accurate enough -- to attempt 
to optimize the execution strategy. For example, splitting large orders into a stream of smaller ones over
time is one of the universally accepted ways to mitigate transaction cost. However, as always, the devil is in the details, and the precise
nature and time scheduling of the orders can make a large difference to the final result.

Optimal execution problems are widespread in the literature and many
different formulas and techniques have been developed to solve
them. However, most of these problems are 
restricted to single asset execution. Recently in \citet{benzquen2016dissecting} we have shown
that even a simple linear model of cross-asset price impact leads to a
very rich phenomenology, in line with the results of~\citet{wang2015price,wang2016microscopic}. In the present paper we provide a practical recipe to
optimize the execution of a portfolio of trades, taking into account
the cross-impact on the different underlying products within the
multivariate framework of~\citet{schied2010optimal}. We will show that proper
synchronization of the legs of the execution schedule is
very important. To quantify the slippage incurred by the strategy we
introduce the EigenLiquidity Model (ELM). The model is
directly related to statistical risk factors which have been used for
portfolio risk management for several decades. Based on a Principal
Component Analysis of the correlation matrix, which provides a
practical method to quantify the amount of different kinds of market
risk (long the market, sectorial, etc.) one can trade while staying
within a prescribed budget of transaction cost.

\section{A quadratic cost model for a single stock}

To set the stage, let us first discuss the execution of a single company's stock over a trading day
that starts at time $t=0$ and lasts until $t=T$. The total volume we have to trade is $V$ shares, which is obtained over the day
by a continuous execution schedule whose local speed is $v(t)$,
normalized as $\int_0^T v(t) \dd t = V$. If we were to hold
this position, our expected risk, defined as the standard deviation of
the daily PnL, would be $\risk = \sigma V$, where $\sigma =
\avg[(p_T-p_0)^2]^{1/2}$ is the daily volatility of the
stock, expressed in dollars. For convenience let us define the trading speed in risk units
as $q(t) := \sigma v(t)$, and the total risk exchanged over the day as
$Q = \int_0^T \dd t \, q(t) := \sigma V$, which naturally equals
$\risk$. 

A standard model for estimating the cost incurred when
trading a certain volume has first been introduced
by~\citet{almgren2001optimal}.
We will consider their framework in a setting in which the trader is
risk-neutral, and impact is linear and
transient~\citep{gatheral2012transient,gatheral2013dynamical,busseti2012calibration,
alfonsi2013capacitary}. This allows one to express the trading costs as
\begin{equation}
  \label{eq:cost_single_stock}
  \cost = \iint_0^T \dd t \, \dd t' q(t) G(t-t') q(t') \, ,
\end{equation}
where $G(\tau)$ is an \emph{impact kernel}~\citep{bouchaud2004fluctuations}, which  quantifies the effect of a small trade $q(t') \dd t'$ on
the price at a later time $t = t'+\tau$. $G(\tau)$ is typically a
decreasing function, dropping from a maximum value obtained at
$\tau=0$ to zero after a slow decay. Consistently with the
results of~\citet{bouchaud2004fluctuations}, it can be written as $G(\tau) = g \phi(\tau)$, with
\begin{equation}
  \label{eq:kernel}
  \phi(\tau) = \left\{
    \begin{array}{lcr}
     (1+\tau/\tau_0)^{-\alpha} 
      &\textrm{for} & \tau \geq 0 \\
      0
      & \textrm{for} & \tau < 0
    \end{array} \right. \, .
\end{equation}
$G(\tau)$ has units of $
1/\$ $, and its inverse corresponds to the amount of risk one would
have to trade, in the absence of decay, to move the stock's price
by its typical daily volatility $\sigma$.

In spite of some limitations (e.g. impact is empirically found to be a
sub-linear function of volume), the model above provides a reasonable
estimation of trading costs for large trades and it is able to
capture the main effects of the trade schedule on costs~\citep{gatheral2012transient,gatheral2013dynamical,
alfonsi2013capacitary}. Extensions of this model to account for risk-aversion have been
considered in~\citep{almgren2001optimal,obizhaeva2013optimal,curato2016optimal}, and for bid-ask spread effects
in~\citep{obizhaeva2013optimal,curato2016optimal}.

\section{Optimal trading of a single stock}
\label{sec:optim-trad-with}

\begin{figure*}
    \centering
    \includegraphics[width=.65\textwidth]{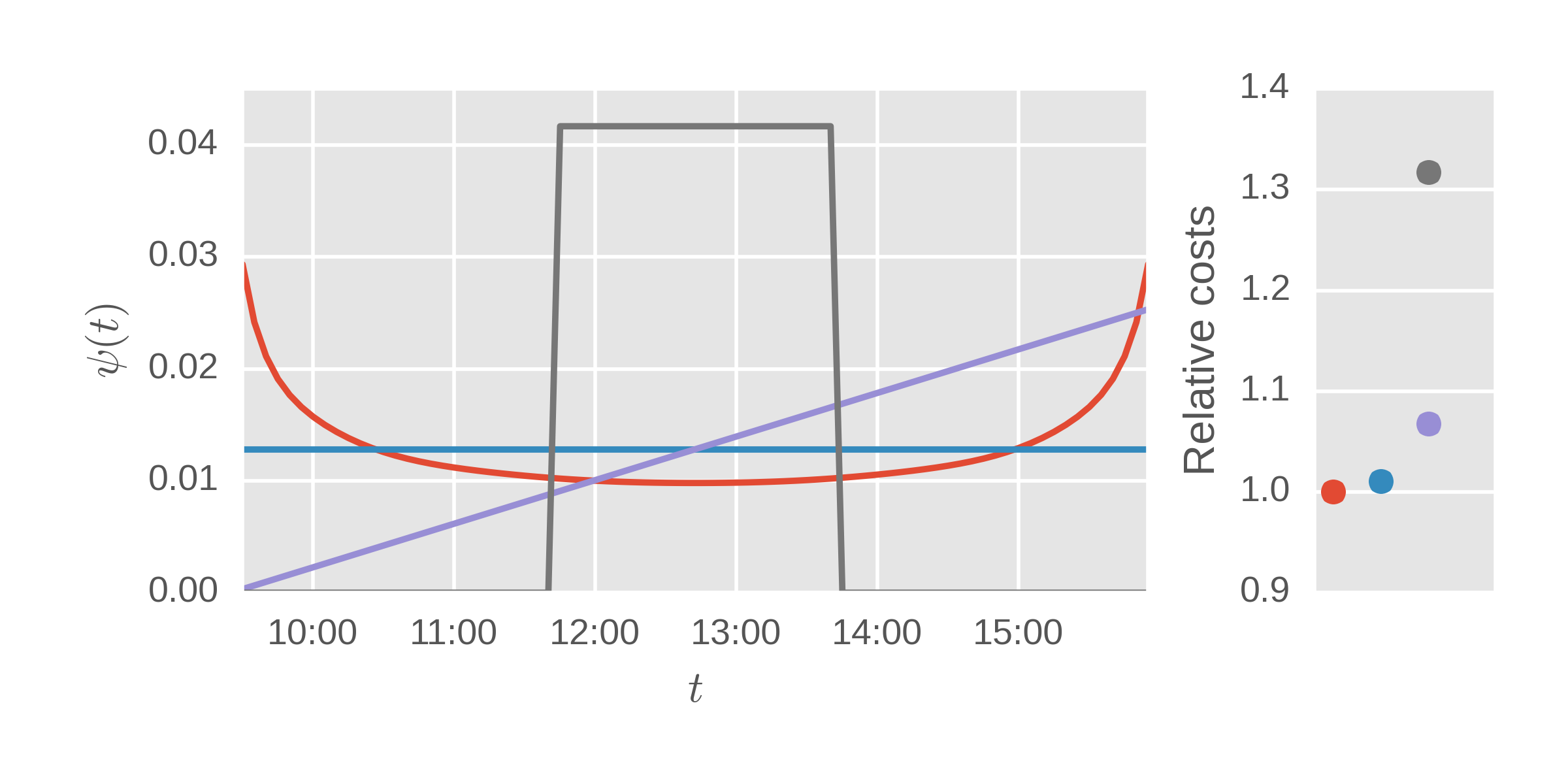}
    \caption{Typical daily trading profiles for executing one unit of daily
      risk (left panel). The red trajectory corresponds to the
      \emph{bucket-shape} trajectory minimizing the trading costs
      under the model in Eq.~\eqref{eq:kernel} with $\alpha = 0.2$ and
      $\tau_0 = 90$ seconds. The blue trajectory is a flat execution rate throughout the day. 
      The grey trajectory corresponds to a flat rate during two hours at mid-day, and the purple trajectory is 
      a linear profile, increasing from morning to afternoon. The
      right panel provides the relative cost of each of the trading
      trajectories with respect to the optimal one.}
    \label{fig:bucket}
  \end{figure*}

Optimal trading schedules under the cost function~\eqref{eq:cost_single_stock} for a fixed
total volume to execute have been extensively investigated in the
finance literature \citep{gatheral2012transient,gatheral2013dynamical,
alfonsi2013capacitary,obizhaeva2013optimal,busseti2012calibration}.
The trajectory of minimum cost can be written as $q(t) = Q
  \psi^\star(t)$, where $\int \dd t \, \psi^\star(t) = 1$ and $\psi^\star(t)$ can be
  determined by solving a linear integral equation~\citep{gatheral2012transient}. This yields
  the well-known symmetric \emph{bucket shape} solution for
  $\psi^\star(t)$ that is depicted in
  Figure~\ref{fig:bucket} (red curve).
  The optimal solution indicates that after an initial
  period of faster trading, one should slow down the execution in order 
  to limit the extra cost due the impact of one's own trades, and then accelerate again the
  trading near the market close. Since trading does not extend beyond that point, 
  strongly impacting the price in this final period does not penalize
  any further executions. 
  As an example, the optimal policy is about 30\% less expensive
  than a localized flat two-hour execution, and approximately 7\% cheaper than
  the linear trading profile represented in Figure~\ref{fig:bucket}.

The temporal shape of the impact kernel precludes \emph{price
    manipulation}, meaning that no round-trip
  trajectory is capable of making money on average~\citep{gatheral2010no,alfonsi2010manipulations,alfonsi2012order}.

\section{A quadratic cost model for portfolios}
\label{sec:cost-impact}
The problem of optimal execution across multiple instruments has been
first considered
in~\citet{schied2010optimal,kratz2014optimal,schoneborn2016adaptive}, whereas the
cost model defined above has been generalized to the
multivariate case in~\citet{alfonsi2016multivariate}
(see also~\citet{schneider2016cross} for a non-linear generalization).
Within that framework, our definition of risk above is readily extended to a portfolio of
multiple stocks with risk positions $\bQ = \{Q^i\}_{i=1}^N$ as
\begin{equation}
  \risk^2 = \bQ^\top \mRho \, \bQ \, ,
\end{equation}
where the correlation matrix $\mRho$ is constructed from the price covariance
matrix $\mSgm = \avg[(\bp_T - \bp_0)(\bp_T - \bp_0)^\top]$ through
\begin{equation}
  \rho^{ij} = \frac{\Sigma^{ij}}{\sqrt{\Sigma^{ii} \Sigma^{jj}}} \, .
\end{equation}
The daily volatilities are generalized to $\bsgm = \{\sigma^i\}_{i=1}^N = \{
(\Sigma^{ii})^{1/2}\}_{i=1}^N$.

Similarly, Eq. \eqref{eq:cost_single_stock} can be extended to this setting as
\begin{equation}
  \label{eq:cost}
    \cost = \iint_0^T \dd t \dd t' \bq^\top(t) \mG (t-t') \bq(t') \, .
\end{equation}

The interpretation of the matrix elements of $\mG(t-t') = \{ G^{ij}(t-t')\}_{ij=1}^N$ is
as before: After trading $\dd q^j(t')$ dollars of risk on the contract $j$ at
time $t'$, we expect the price of contract $i$ to change by
$G^{ij}(t-t') \dd q^j(t')$ units of its daily dollar volatility $\sigma^i$. 
The terms with $i=j$ correspond to direct price impact, which was already described by earlier models where each stock is independent. 
In addition, the new terms with $i\neq j$ describe \emph{cross-impact} between stocks which, as it was shown by \citet{benzquen2016dissecting}, 
is a highly relevant effect since it explains an important fraction of the cross-correlation between stocks, a feature that we will use below.

\citet{benzquen2016dissecting} have further found that within a
good degree of approximation, one can write the kernel
$\mG(\tau)$ in the factorized form $\mG(\tau) =
\left[\mG_+ + \mG_-\right] \phi(\tau)$, with $\phi(\tau)$ given by Eq.~\eqref{eq:kernel}, and
$\mG_\pm$ denote respectively the symmetric and antisymmetric part
of $\mG$. \citet{schneider2016cross} have shown that when the antisymmetric
part of the propagator $\mG_-$ is large, then price manipulation is possible, leading to an ill-defined
cost of optimal strategies. Since empirically $\mG_-$ is small, we set it to zero so that the cost
associated with the execution of a portfolio of trades reads:
\begin{equation}
  \label{eq:symcost}
    \cost = \frac{1}{2}
    \iint_0^T \dd t \dd t' \left( \bq^\top(t) \mG_+ \bq(t') \right) \phi(|t-t'|)
    \, .
\end{equation}
The condition of symmetry for $\mG$ is not the only one required in
order not to have price manipulation. In fact,
formula~\eqref{eq:symcost} is free of price manipulation if and only
if $\mG_+$ is positive semidefinite.
This amounts to saying that buying a
portfolio always pushes its price up and vice versa, regardless of its
composition, resulting in an impact cost that is always greater or equal
to zero independently of the portfolio that is actually traded.

\section{The EigenLiquidity Model}

In principle, $\mG$ can be determined using simultaneous Trades and Quotes data
for the corresponding pool of stocks. However, the
empirical impact matrix $\mG$ is in general extremely noisy,
so some cleaning scheme is necessary.
By leveraging the empirical results of \citet{benzquen2016dissecting} (see Fig.7
therein), showing that the structure of the impact matrix is in a
suitable statistical sense ``close'' to the one of the correlation
matrix, we make the assumption  that the
impact matrix has the {\it same  set of eigenvectors} as those of the correlation matrix
$\mRho$. Intuitively, the eigenvectors of $\mRho$ correspond to portfolios with uncorrelated returns. Our 
assumption means, quite naturally, that trading one of these portfolios will only impact (to first approximation) the
returns of that portfolio, but not of any other orthogonal
one. Besides being an empirically reasonable cleaning scheme for $\mG$, this
choice is motivated by the results illustrated in this section, showing that
it leads to a cost function $\cost$ that satisfy three fundamental
consistency requirements: symmetry, positive semi-definiteness and
fragmentation invariance.

More precisely, one can write
\begin{eqnarray}
  \label{eq:rho}
  \mRho = \mO \mLam \mO^\top \, ,
\end{eqnarray}
where $\mO = \{O^{ia}\}_{ia=1}^N$ is an $N\times N$ orthogonal matrix
of eigenvectors and $\mLam = \{\Lambda^a \delta^{ab}\}_{ab=1}^N$ is a diagonal
matrix of $N$ non-negative eigenvalues. Our assumption is that the matrix $\mG$ has the following structure:
\begin{equation}
  \label{eq:film_prop}
  \mG = \mO \mLam \mLiq \mO^\top \,:= \mRho^{1/2} \mLiq (\mRho^{1/2})^\top\, ,
\end{equation}
where $\mLiq=\{\liq^a \delta^{ab}\}_{ab=1}^N$ is a diagonal matrix, and $\mRho^{1/2} = \mO
\mLam^{1/2}$.

An important property of this decomposition is that it leads to a \emph{fragmentation invariant}
cost formula in the following sense: When trading two completely
correlated products $i$ and $j$ (i.e., when $\rho^{ij} = 1$), the impact of a trading trajectory does not
depend on how the volume is split between the two instruments.
More formally, a fragmentation invariant cost $\cost$ is left unchanged under the transformation
\begin{eqnarray}
  q^{i}(t) &\to& q^i(t) + \delta q(t) \\
  q^{j}(t) &\to& q^j(t) - \delta q(t) \, ,
\end{eqnarray}
where $\delta q(t)$ is completely arbitrary. Intuitively, Eq.~\eqref{eq:film_prop} fulfills this property
because of the factor $\mLam$ multiplying $\mLiq$. If instruments $i$ and $j$ are completely correlated, then the
relative mode ``rel'' is an eigenvector of zero risk, with $\Lambda^{\textrm{rel}} = 0$. When used for estimating the
cost of an execution trajectory $\bq(t)$, Eq.~\eqref{eq:film_prop}
will single out such relative modes through the projection $\mO^\top
\bq(t)$, and it will weight them by the corresponding risk
$\mLam$ which is zero. 

The impact model~\eqref{eq:film_prop} will be called the
EigenLiquidity Model (ELM). It can be seen as the most natural choice among all the models 
implementing fragmentation invariance. In fact, it
\emph{continuously} interpolates between small
risk modes (that are thus expected to be characterized by small impact) and large risk modes 
(for which impact costs can be substantial).

Empirically, Eq.~\eqref{eq:film_prop} has been shown to hold to a
good degree of approximation in~\citet{benzquen2016dissecting}.
On the other hand, the condition of positive semi-definiteness of $\mG$, i.e. $\liq^a \geq 0$, $\forall a$,
is not guaranteed from Eq.~\eqref{eq:film_prop} and thus should be checked
on empirical data. This is what we display in
Figure~\ref{fig:eig_g} using real data, confirming that all the
$\liq^a$'s are actually strictly positive.
The quantity $(\liq^a)^{-1}$ has the natural interpretation of a
\emph{liquidity per mode}. It expresses the amount of daily risk in
dollars that one can trade on the eigen-portfolio $a$ to move its price
by its daily volatility $\sqrt{\Lambda^a}$.

\section{Optimal trading of portfolios}

Under the ELM, the impact cost of any schedule
$\bq(t)$ admits an interpretation in terms
of the modes of the correlation matrix of normalized returns through
the decomposition
  \begin{equation}
    \label{eq:eigencost}
    \cost = \frac 1 2 \sum_{a=1}^N g^a ||\widetilde q^a||^2\, ,
  \end{equation}
where $|| \widetilde q^a ||^2 = \iint_0^T \dd t \dd t' \, \widetilde q^a(t)
\phi(|t-t'|) \widetilde q^a(t')$.
We have also introduced the notation
\begin{equation}
  \widetilde \bq(t) = (\mRho^{1/2})^\top \bq(t)
  \label{eq:qa}
\end{equation}
denoting the projection of the executed volumes on a set of
\emph{uncorrelated, unit risk eigen-portfolios} $\mRho^{-1/2} = \{\bpi^a \}_{a=1}^N$.

The notion of eigen-portfolios is very useful for intuitively characterizing
the cost formula~\eqref{eq:eigencost}. The name
comes from the fact that these portfolios are not only uncorrelated
and unit-risk (i.e., $(\bpi^a)^\top \mRho \bpi^b =
\delta^{ab}$), but furthermore trading an amount of the basket $a$
according to the weights given by $\bpi^a$ has no impact on the total
value of the basket $b$ and vice versa (i.e., $(\bpi^a)^\top \mG \bpi^b =
\delta^{ab}\liq^a $). This is precisely the intuition behind our central assumption, Eq.~\eqref{eq:film_prop}.

This construction implies that the cost $\cost$ can be calculated by
first projecting the strategy $\bq(t)$ on the portfolios $\bpi^a$ via Eq.~\eqref{eq:qa} and then by
taking a sum of an impact cost per  
mode $\liq^a$ with a weighting factor $||\widetilde q^a||^2$ given by such projections.

Eq.~\eqref{eq:eigencost} also shows clearly that the
positivity of the matrix $\mLiq$ and the kernel $\phi(\tau)$ makes the
optimization problem convex, which always has a unique solution. The optimum under the terminal
constraint $\bQ = \int_0^T \dd t \, \bq(t)$ is necessarily
achieved under a \emph{synchronous} execution schedule where at any
given point in time all stocks are traded with the same time profile,
i.e., $\bq(t) = \bQ \psi^\star(t)$, resembling the case without
cross-impact.

Intuitively, an asynchronous execution strategy can be seen in the mode space
as the optimal one above, plus a round-trip along some of these modes.
The convexity of the cost function~\eqref{eq:symcost} implies that round-trips
always {\it increase} execution costs, so they should be avoided.\footnote{One may compare this with a related discussion by \citet{wang2017trading} investigating the case of round-trips on two stocks.}
Hence, synchronicity is a general consequence
of the convexity of the problem, together with the homogeneity of the
decay kernels $\phi(\tau)$ for different instruments.

A toy example to explain the
implications of the formalism is given in Box \ref{box:1}.

\begin{table}
\begin{tcolorbox}
\textbf{Box \ref{box:1}. A toy example with two stocks}

A simple realistic case is $N=2$ stocks and an impact matrix $\mG$ of the
form
\begin{boxoneequation}
  \mG = \left(
    \begin{array}{cc}
      G^{\textrm{diag}} & G^{\textrm{off}} \\
      G^{\textrm{off}}  & G^{\textrm{diag}} 
    \end{array}
\right) \, ,
\end{boxoneequation}
\noindent with $G^{\textrm{diag}} > G^{\textrm{off}}$.
Let us also suppose the target volumes to have the same magnitude:
\begin{boxoneequation}
  \label{eq:4}
  \bQ = (Q^1,Q^2) = (Q,\pm Q) \, .
\end{boxoneequation}
After defining $\psi^{(i)}(t) = q^i(t) / Q^i$, the cost becomes
\begin{boxoneeqnarray}{1}
  \cost = \frac {|Q|^2} 4 \left[
  \underbrace{(G^{\textrm{diag}} + G^{\textrm{off}})}_{G^\abs} || \psi^{(1)} \pm
  \psi^{(2)}||^2 + \right. \nonumber \\ \left.
  \underbrace{(G^{\textrm{diag}} - G^{\textrm{off}})}_{G^\rel} || \psi^{(1)} \mp \psi^{(2)}||^2
  \right]
\end{boxoneeqnarray}

The interpretation of this result is the following:
\begin{itemize}
\item The cost of trading is proportional to the eigenvalues $G^\abs$ and
  $G^\rel$ (where $\abs$ and $\rel$ stand for \emph{absolute} and
  \emph{relative} mode). It is obviously minimized by choosing $\psi^{(1)} =
  \psi^{(2)} = \psi^\star$, in which case the cost is $|Q|^2 G^{\abs/\rel}
  ||\psi^\star||^2$
\item When trading directionally (i.e., if $Q^1 = Q^2$), the minimum cost is
  proportional to $G^\abs$, while the neutral strategy $Q^1 = -Q^2$ 
  yields a smaller cost proportional to $G^\rel$.
\item One could be tempted to locally trade the cheaper relative mode, but
  this would construct a long-short position which would have to be closed at a
  cost later. It is easy to check that the convexity of the cost and the terminal
  requirement prevent this from being optimal.
\end{itemize}
Which trajectory is cheaper
in terms of risk? If we assume the correlation matrix to be given by \\
$
  \mRho = \left(
    \begin{smallmatrix}
      1 & \rho \\
      \rho  & 1 
    \end{smallmatrix}
  \right) \, ,
$
\noindent its eigenvalues are equal to $\Lambda^{\abs/\rel} = 1 \pm \rho$. The cost of
trading per unit of risk can be written as
\begin{boxoneequation}
  \frac {\cost}{\risk} 
= ||\psi^\star||^2 \frac{|Q| G^{\abs/\rel}}{\sqrt{2\Lambda^{\abs/\rel}}}
= ||\psi^\star||^2 \frac{|Q| g^{\abs/\rel}\sqrt{\Lambda^{\abs/\rel}}}{\sqrt{2}} \, ,
\end{boxoneequation}
\noindent where $(g^{\abs/\rel})^{-1}$ is the liquidity in dollars of
the absolute and relative mode, respectively.
We can interpret this as follows:
\begin{itemize}
\item The cost of trading per unit risk depends trivially on $G^{\abs/\rel}$,
  but it has an implicit dependence on the correlation through
  $\Lambda^{\abs/\rel}$. The more correlated are the stocks, the more
  expensive it is to obtain a target risk.
\item The liquidity per mode $(g^{\abs/\rel})^{-1}$ accounts for both effects,
  describing how expensive it is to obtain a given target risk by
  trading either the symmetric or the antisymmetric mode.
\end{itemize}
\refstepcounter{table}
\label{box:1}
\end{tcolorbox}
\end{table}

\begin{table}[!t]
\begin{tcolorbox}
\textbf{Box \ref{box:2}. Fitting the EigenLiquidity Model}

We present a step-by-step procedure for calibrating our cost model to real data.

\begin{enumerate}
    \item Compute the covariance matrix of prices
      \begin{boxtwoequation}
        \mSgm = \avg[(\bp_{T} - \bp_0)(\bp_{T} - \bp_{0})^\top],
      \end{boxtwoequation}
      and extract the volatilities $\sigma^i$.
    \item Standardize prices, their covariances, and market volumes:
      \begin{boxtwoeqnarray}{3}
        x^i_t & = & p^i_t / \sigma^i, \\
        \refstepcounter{boxtwocounter}
        \rho^{ij} &=& \Sigma^{ij} / (\sigma_i\sigma_j), \\
        \refstepcounter{boxtwocounter}
        q^i_t & = & \sigma^i v^i_t.
      \end{boxtwoeqnarray}
    \item Compute the covariation of prices and volumes:
      \begin{boxtwoeqnarray}{3}
        \refstepcounter{boxtwocounter}
        \mR(t-t') &=& \avg[\dot \bx_t \bq_{t'}^\top], \\
        \refstepcounter{boxtwocounter}
        \mC(t-t') &=& \avg[\bq_t \bq_{t'}^\top],
      \end{boxtwoeqnarray}
      where $\dot \bx_t = (\bx_{t+\dd t} - \bx_t) / \dd t$.
    \item Compute the derivative of the kernel $\dphi(\tau) = [\phi(\tau + \dd
      \tau) - \phi(\tau)] / \dd \tau$ by solving
      \begin{boxtwoequation}
        \bar r(t-t') \propto \int_{-\infty}^t \dd t'' \dphi(t-t'') \bar c(t''-t'),
      \end{boxtwoequation}
      where $\bar r(\tau) = N^{-1} \sum_{i=1}^N r^{ii}(\tau)$ and
      $\bar c(\tau) = N^{-1} \sum_{i=1}^Nc^{ii}(\tau) /
      c^{ii}(0)$. Get the norm from the condition $\phi(0) = 1$.
    \item Extract indepedent portfolios $\{\bpi^a \}_{a=1}^N$ from
      the correlation matrix $\mRho$ by computing eigenvectors $\mO$ and
      eigenvalues $\mLam$ according to Eq.~\eqref{eq:rho}.

    \item Project the covariances $\mR(\tau)$ and $\mC(\tau)$ to indepedent
      portfolios:
      \begin{boxtwoeqnarray}{2}
        \label{eq:12}
        \widetilde r^a(\tau) &=& (\bpi^a)^\top \mR(\tau) \bpi^a, \\
        \refstepcounter{boxtwocounter}
        \widetilde c^a(\tau) &=& (\bpi^a)^\top \mC(\tau) \bpi^a.
      \end{boxtwoeqnarray}
    \item Estimate $g^a$ with the maximum-likelihood estimator
      \begin{boxtwoequation}
        \label{eq:11}
        g^a = \frac{1}{\Lambda^a} \left(
          \frac{\int_0^\infty \dd \tau \, \dphi(\tau)
          \widetilde r^a(\tau)}
        {\iint_{-\infty}^{+\infty} \dd t \dd t' \, \dphi(t)
          \dphi(t') \widetilde c^a(t-t')} \right).
      \end{boxtwoequation}
\end{enumerate}
\end{tcolorbox}
\refstepcounter{table}
\label{box:2}
\end{table}

\section{Applications to real data}
\label{sec:applications}

We have fitted the ELM to a pool of 150 US stocks in 2012, following
the procedure described in detail in
Box~\ref{box:2}. In~\citet{benzquen2016dissecting} 
for the time decay of impact we find $\alpha
\approx 0.15$, and $\tau_0 \approx 90$ seconds.

\begin{figure*}
  \centering
  \includegraphics[width=.6\textwidth]{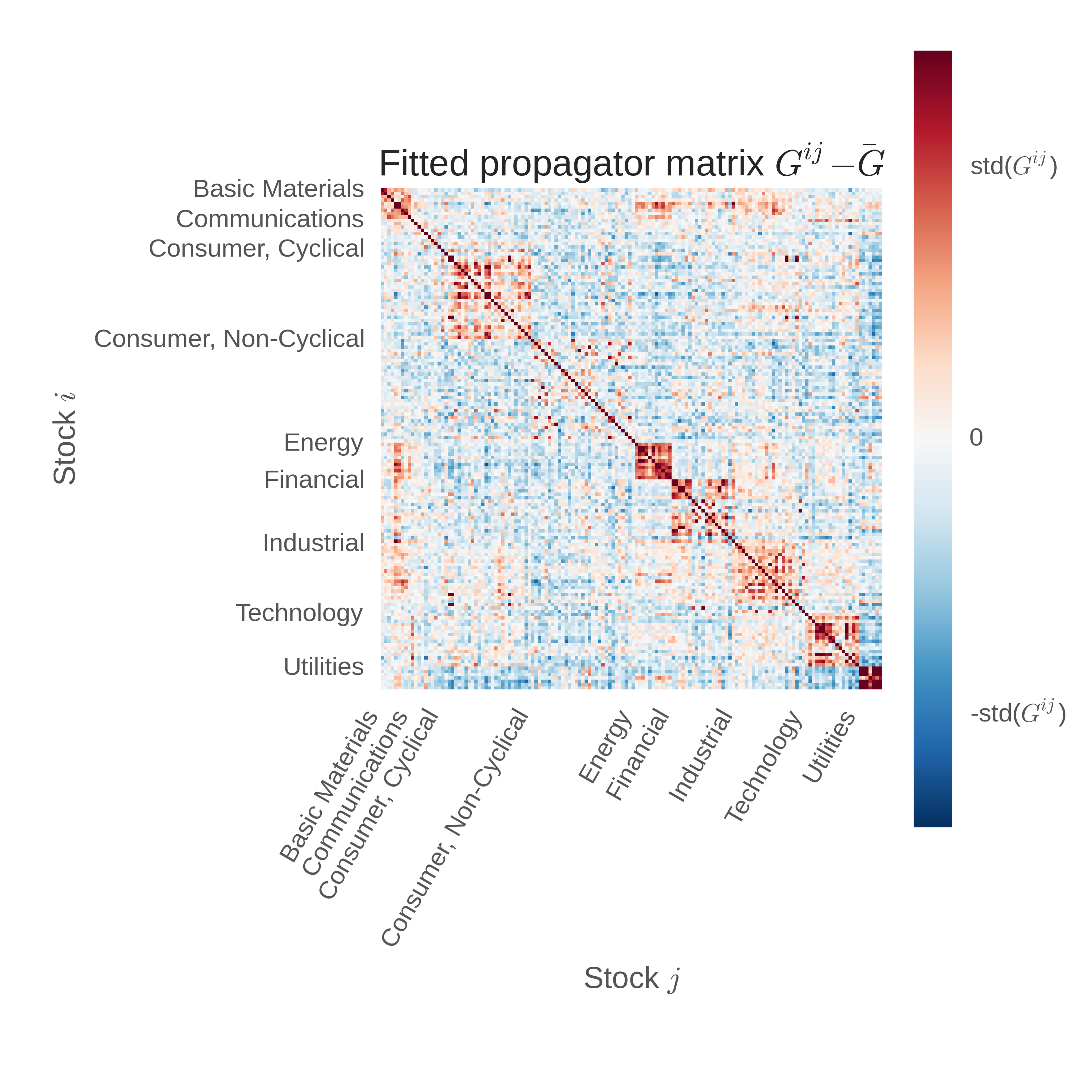}
  \caption{Propagator $\mG$ fitted during the year 2012 on a sample of
    150 US stocks sorted 
    by industrial sectors. The market mode (i.e., the average across
    the entries $\bar G$) has been removed in order to highlight the
    sectorial structure of the market. The figure clearly shows that
    after hiding the market mode --~that accounts for the overall
    positive interaction between buy (sell) trades and positive
    (negative) price changes~-- the other large modes of the
    propagator matrix can be interpreted  as financial sectors, which
    are responsible for the  block structure of the matrix $\mG$.}
  \label{fig:heatmap}
\end{figure*}

A visual representation of the impact matrix is shown in Figure~\ref{fig:heatmap}. The inhomogeneity of $\mG$ captures the sectorial
structure of the market, encoding the specific dependence of stock $i$ on its sector and/or its most correlated stocks $j$.

\begin{figure*}
  \centering
  \includegraphics[width=.6\textwidth]{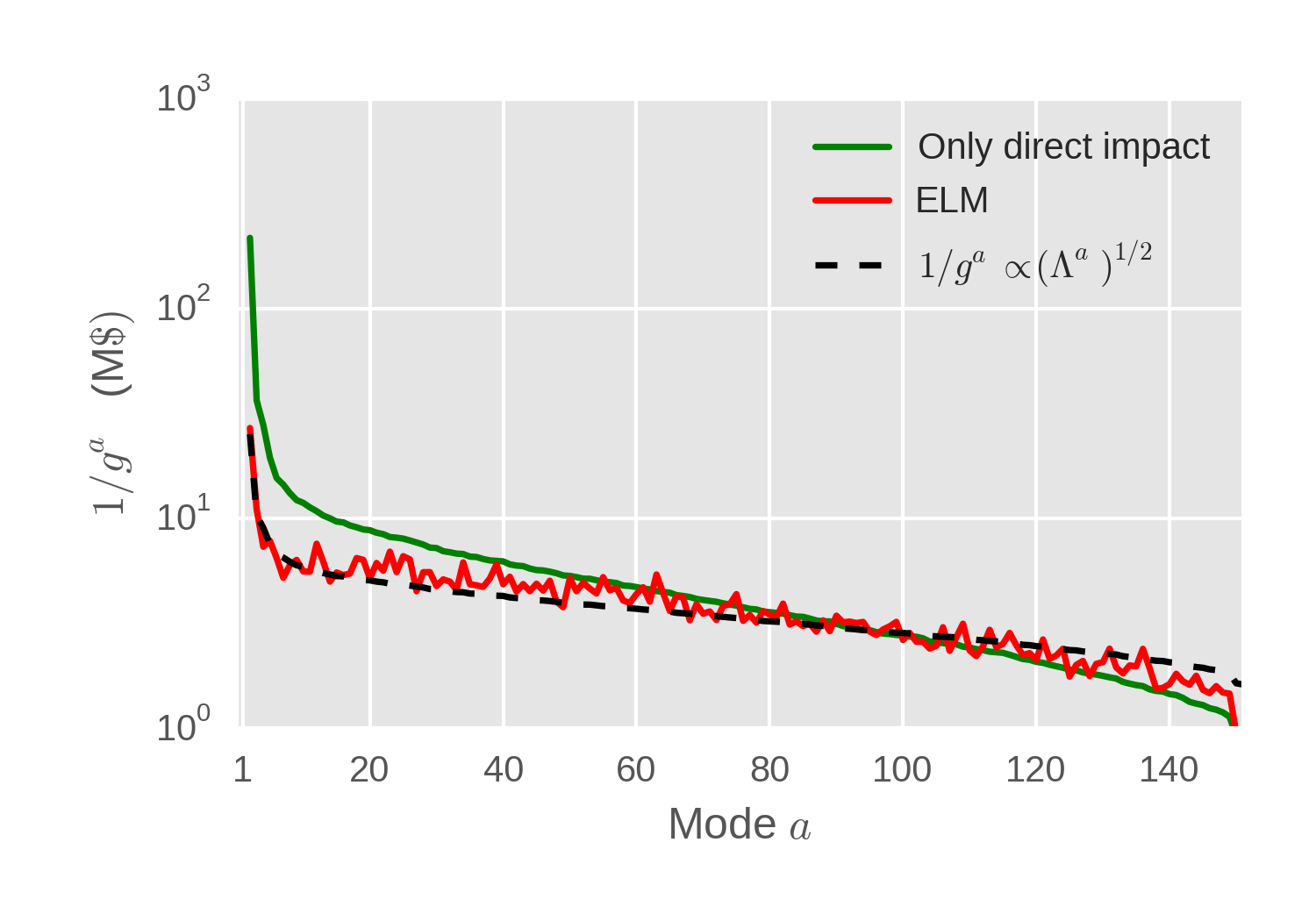}
  \caption{Liquidity per mode $(g^a)^{-1}$, obtained by normalizing
    the cost of trading the portfolio $\bpi^a$ by its corresponding risk (red
    line). Each eigenvalue $g^a$ is interpreted as the cost of trading a dollar of risk
    in the portfolio $\bpi^a$, so its inverse gives the liquidity in
    dollars available
    on the mode $a$. The green line represents the liquidity available
     $\bpi^a$ under a model in which no
    cross-interaction is taken into account, indicating that when
    neglecting cross-impact one underestimates the cost of trading
    high-risk modes (e.g., the market) and overestimates the cost of
    trading the small-risk ones. The dashed line plotted for
    comparison indicates the
    prediction of a model in which $g^a \propto (\Lambda^a)^{-1/2}$.
    Note that while the liquidity  of the large-risk
    modes (left part of the plot) is relatively easy to estimate from
    empirical data, a preliminary cleaning step on the noisy part of
    the eigenvalue spectrum is further required in order to remove the
    spectrum of the bulk of lower-risk modes on the right side of the
    plot~\citep{bun2016cleaning}.}
  \label{fig:eig_g}
\end{figure*}

The main difference of our ELM and a model without cross-impact is that
the impact costs of a large buy program can no longer be reduced as effectively by spreading the orders 
across multiple correlated instruments. The impact kernel diffuses the interaction across markets and sectors
through the modes. The transaction cost of trading $Q$ dollars of
risk in the mode $\bpi^a$ is equal to $g^a Q^2$ dollars. Figure~\ref{fig:eig_g} shows the inverse of the eigenvalues
$g^a$, this is about 30M dollars of risk on the most liquid
mode. Most of the cross-interaction between stocks is captured by this
\emph{market mode}, which accounts for the fact that when buying a
dollar of risk of a stock picked at random in our pool, the other
stocks in the pool rise on  average by $\bar G \approx 0.4 \times 10^{-4}$ times their daily
volatility, where $\bar G$ is the average of the off-diagonal elements of $\mG$. 
Smaller risk modes may be up to 30 times less liquid.
Empirically, the liquidity per mode $g^a$ is well fitted by $g^a \propto
(\Lambda^a)^{-1/2}$, see Figure~\ref{fig:eig_g}. This finding is consistent with the assumption of
fragmentation invariance, which implicitly requires the parameters
$g^a \Lambda^a \to 0$ when $\Lambda^a \to 0$ (see Box~\ref{box:2}).  

In order to illustrate the relevance of these findings when executing
a portfolio of trades, let us study numerically a toy daily execution problem of a 
trader who has target volumes $\bQ$ corresponding to a fraction
$\varphi=$ 1\%,5\%, 10\% of the daily
liquidity of each of $N =150$ US stocks. We assume that the trader uses the optimal,
synchronous policy derived above: $\bpsi(t) = \bpsi^\star(t)$.

To explore a variety of trading styles, we set the sign ($\epsilon=+$ for buy,
$\epsilon=-$ for sell) of the $N$ orders from $N$ biased coin tosses. We vary the bias parameter $\beta =
\avg[\epsilon^i]$ in the interval $[-1,+1]$. The interpretation of this construction for $\beta=0$ or $\pm 1$ is very simple:
\begin{itemize}
\item For $\beta=\pm1$, the order is a long or short directional one, and is
  strongly exposed to the market mode of risk.
\item For $\beta = 0$, the strategy is neutral, and its
  exposition to the market mode is therefore limited.
\end{itemize}

The average cost under such an execution policy can be expressed
analytically, allowing to obtain a relation between the cost $\cost$ and $\beta$:

\begin{eqnarray}
  {\avg[\cost]} =
  \frac{\varphi^2 ||\psi^\star||^2}{2} \left[ {{ (1-\beta^2) \sum_i
  G^{ii} (Q^i_{\textrm{M}})^2 + \beta^2 \bQ^\top_{\textrm{M}}\mG
  \bQ_{\textrm{M}} }} \right], \label{eq:cost_turn}
\end{eqnarray}
where $\bQ_{\textrm{M}}$ denotes average dollar risk exchanged by the
market on the different stocks.

\begin{figure*}
  \centering
  \includegraphics[width=.5\textwidth]{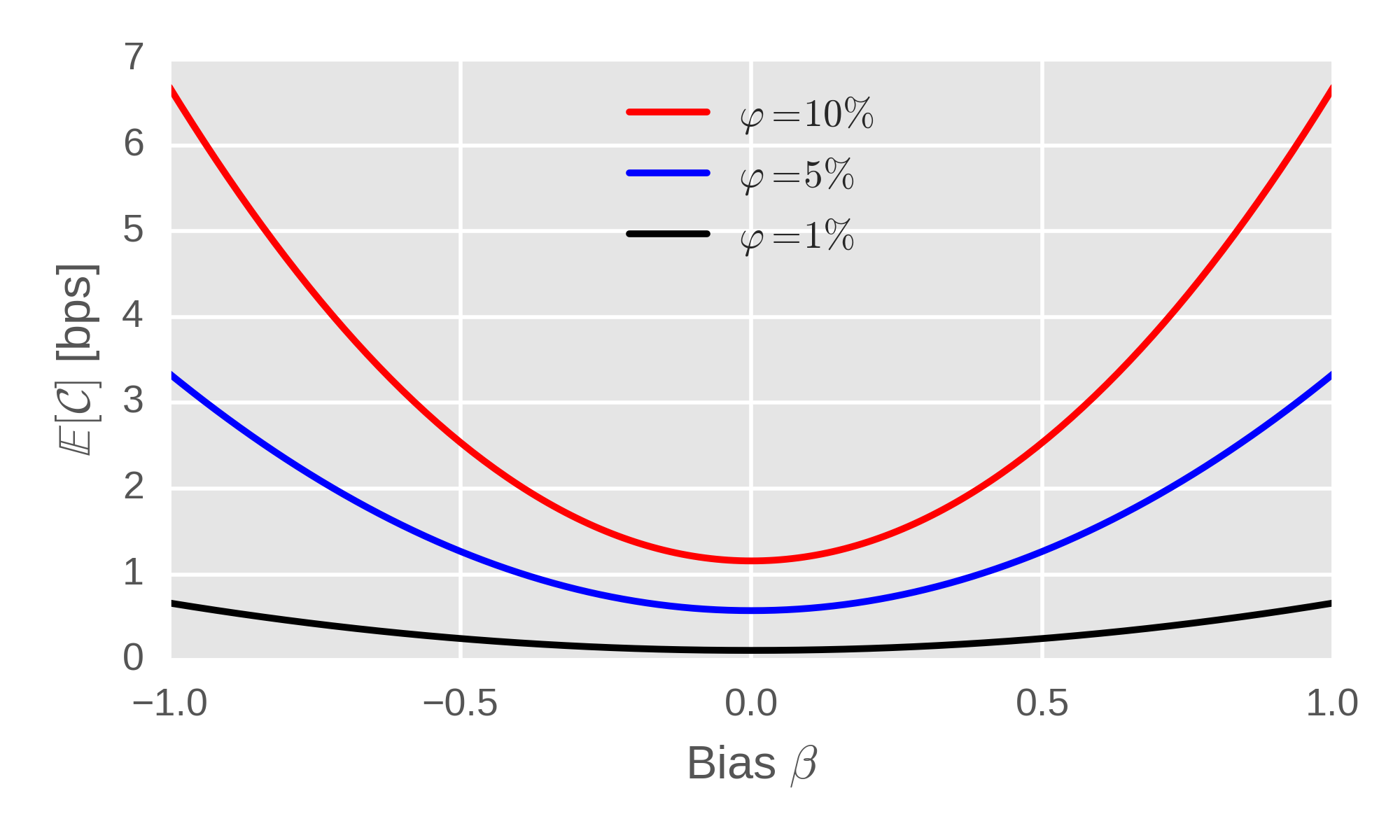}
  \caption{Average trading cost $\cost$ in basis points for the
    trading strategy described in Section~\ref{sec:applications}, as a
    function of the bias parameter $\beta$, for different participation rates
    of $\varphi$ =  1\%, 5\%, 10\%. Consistently with Eq.~\eqref{eq:cost_turn}, one
    can clearly see that directional trading strategies are more expensive
    in terms of notional traded.}
  \label{fig:bias}
\end{figure*}

Comparing Eq.~\eqref{eq:cost_turn} between the above special cases, one can see 
that the dollar cost is higher for $\beta=\pm1$ than for $\beta=0$ by a factor that can be
estimated on the basis of Eq.~\eqref{eq:cost_turn} to roughly be $g^1 \Lambda^1 / ( N^{-1} \sum_a g^a
\Lambda^a ) \approx 6.6$. This makes sense, as this is the ratio of the top eigenvalue of
$\mG$, expressing the cost of trading the market directionally, versus the
average of all eigenvalues selecting the direct impact contribution in the left
term of Eq.~\eqref{eq:cost_turn}. Figure~\ref{fig:bias}
shows the full evolution of this ratio as a function of the bias $\beta$.

However, this result holds for a fixed dollar volume traded, but the risk of the resulting position is in fact 
much higher for $\beta=1$ than for $\beta=0$. The cost of trading per unit risk taken, expressed by
$\avg[\cost]/\sqrt{\avg[\risk^2]}$ is found to be approximately
independent of $\beta$. By generalizing Eq.~\eqref{eq:cost_turn} to the
ratio $\avg[\cost] /\sqrt{\avg[\risk^2]}$, one can relate this finding
to the numerical result $g^1 (\Lambda^1)^{1/2} \approx 1.1 \times \sum_a
g^a \Lambda^a / \sqrt{\sum_a \Lambda^a}$.

\section{Conclusions}
\label{sec:conclusions}
In this paper we have shown how to leverage the recent quantitative results of
\citet{benzquen2016dissecting} on cross-impact effects in order to estimate the execution
cost of a basket of correlated instruments.
We confirm empirically on a pool of 150 US stocks that cross-impact is a
very substantial part of the impact of  trades on prices. We show that
neglecting cross-interactions leads to a distorted vision of the
liquidity available on the market: it overestimates the liquidity on
large risk modes and underestimates the liquidity of low-risk modes.

In order to distill these findings into a cost formula, we
have assumed that the impact matrix has the same eigenvectors as the correlation matrix
itself, and impact eigenvalues are proportional to the risk of the corresponding modes. 
This specification prevents arbitrage opportunities and price manipulation strategies. 
It also abides by the principle of fragmentation invariance,
which states that trading zero risk portfolios should have no effect
whatsoever on trading costs. We have provided the solution of the corresponding optimal trading
problem, which leads to a synchronous U-shaped trading profile across products. This avoids
round trips on unwanted positions at a potentially large cost.

In order to keep our approach as simple as possible, we have neglected other
sources of cost (spread costs, fees), and considered no risk aversion effects nor intraday predictive signals.
Moreover, we have deliberately disregarded the non-linear
nature of the price impact function, which is known to be better represented by a square-root law \citep{grinold2000active, toth2011anomalous,schneider2016cross}.
These features could be progressively reintroduced into
our framework by preserving the idea of an interaction that is
diagonal in the space of correlation modes. Indeed, we believe our
simper approach to be better suited in order to illustrate
transparently the main effects of cross-impact between financial instruments.

\bibliographystyle{abbrvnat}
\bibliography{bibs}

\begin{thebibliography}{23}
\providecommand{\natexlab}[1]{#1}
\providecommand{\url}[1]{\texttt{#1}}
\expandafter\ifx\csname urlstyle\endcsname\relax
  \providecommand{\doi}[1]{doi: #1}\else
  \providecommand{\doi}{doi: \begingroup \urlstyle{rm}\Url}\fi

\bibitem[Alfonsi and Schied(2010)]{alfonsi2010manipulations}
A.~Alfonsi and A.~Schied.
\newblock Optimal trade execution and absence of price manipulations in limit
  order book models.
\newblock \emph{SIAM Journal on Financial Mathematics}, 1\penalty0
  (1):\penalty0 490--522, 2010.

\bibitem[Alfonsi and Schied(2013)]{alfonsi2013capacitary}
A.~Alfonsi and A.~Schied.
\newblock Capacitary measures for completely monotone kernels via singular
  control.
\newblock \emph{SIAM Journal on Control and Optimization}, 51\penalty0
  (2):\penalty0 1758--1780, 2013.

\bibitem[Alfonsi et~al.(2012)Alfonsi, Schied, and Slynko]{alfonsi2012order}
A.~Alfonsi, A.~Schied, and A.~Slynko.
\newblock Order book resilience, price manipulation, and the positive portfolio
  problem.
\newblock \emph{SIAM Journal on Financial Mathematics}, 3\penalty0
  (1):\penalty0 511--533, 2012.

\bibitem[Alfonsi et~al.(2016)Alfonsi, Kl{\"o}ck, and
  Schied]{alfonsi2016multivariate}
A.~Alfonsi, F.~Kl{\"o}ck, and A.~Schied.
\newblock Multivariate transient price impact and matrix-valued positive
  definite functions.
\newblock \emph{Mathematics of Operations Research}, 41\penalty0 (3):\penalty0
  914--934, 2016.

\bibitem[Almgren and Chriss(2001)]{almgren2001optimal}
R.~Almgren and N.~Chriss.
\newblock Optimal execution of portfolio transactions.
\newblock \emph{Journal of Risk}, 3:\penalty0 5--40, 2001.

\bibitem[Benzaquen et~al.(2017)Benzaquen, Mastromatteo, Eisler, and
  Bouchaud]{benzquen2016dissecting}
M.~Benzaquen, I.~Mastromatteo, Z.~Eisler, and J.-P. Bouchaud.
\newblock Dissecting cross-impact on stock markets: An empirical analysis.
\newblock \emph{Journal of Statistical Mechanics: Theory and Experiment},
  2017\penalty0 (2):\penalty0 023406, 2017.

\bibitem[Bouchaud et~al.(2004)Bouchaud, Gefen, Potters, and
  Wyart]{bouchaud2004fluctuations}
J.-P. Bouchaud, Y.~Gefen, M.~Potters, and M.~Wyart.
\newblock Fluctuations and response in financial markets: the subtle nature of
  ``random" price changes.
\newblock \emph{Quantitative Finance}, 4\penalty0 (2):\penalty0 176--190, 2004.

\bibitem[Bun et~al.(2016)Bun, Bouchaud, and Potters]{bun2016cleaning}
J.~Bun, J.~Bouchaud, and M.~Potters.
\newblock Cleaning correlation matrices.
\newblock \emph{Risk Magazine}, 2016.

\bibitem[Busseti and Lillo(2012)]{busseti2012calibration}
E.~Busseti and F.~Lillo.
\newblock Calibration of optimal execution of financial transactions in the
  presence of transient market impact.
\newblock \emph{Journal of Statistical Mechanics: Theory and Experiment},
  2012\penalty0 (09):\penalty0 P09010, 2012.

\bibitem[Curato et~al.(2016)Curato, Gatheral, and Lillo]{curato2016optimal}
G.~Curato, J.~Gatheral, and F.~Lillo.
\newblock Optimal execution with non-linear transient market impact.
\newblock \emph{Quantitative Finance}, 17\penalty0 (1):\penalty0 41--54, 2016.

\bibitem[Gatheral(2010)]{gatheral2010no}
J.~Gatheral.
\newblock No-dynamic-arbitrage and market impact.
\newblock \emph{Quantitative Finance}, 10\penalty0 (7):\penalty0 749--759,
  2010.

\bibitem[Gatheral and Schied(2013)]{gatheral2013dynamical}
J.~Gatheral and A.~Schied.
\newblock Dynamical models of market impact and algorithms for order execution.
\newblock \emph{Handbook on Systemic Risk, Jean-Pierre Fouque, Joseph A.
  Langsam (eds.)}, pages 579--599, 2013.

\bibitem[Gatheral et~al.(2012)Gatheral, Schied, and
  Slynko]{gatheral2012transient}
J.~Gatheral, A.~Schied, and A.~Slynko.
\newblock Transient linear price impact and fredholm integral equations.
\newblock \emph{Mathematical Finance}, 22\penalty0 (3):\penalty0 445--474,
  2012.

\bibitem[Grinold and Kahn(2000)]{grinold2000active}
R.~C. Grinold and R.~N. Kahn.
\newblock \emph{Active portfolio management}.
\newblock McGraw-Hill New York, NY, 2000.

\bibitem[Kratz and Sch{\"o}neborn(2014)]{kratz2014optimal}
P.~Kratz and T.~Sch{\"o}neborn.
\newblock Optimal liquidation in dark pools.
\newblock \emph{Quantitative Finance}, 14\penalty0 (9):\penalty0 1519--1539,
  2014.

\bibitem[Obizhaeva and Wang(2013)]{obizhaeva2013optimal}
A.~A. Obizhaeva and J.~Wang.
\newblock Optimal trading strategy and supply/demand dynamics.
\newblock \emph{Journal of Financial Markets}, 16\penalty0 (1):\penalty0 1--32,
  2013.

\bibitem[Schied et~al.(2010)Schied, Sch{\"o}neborn, and
  Tehranchi]{schied2010optimal}
A.~Schied, T.~Sch{\"o}neborn, and M.~Tehranchi.
\newblock Optimal basket liquidation for cara investors is deterministic.
\newblock \emph{Applied Mathematical Finance}, 17\penalty0 (6):\penalty0
  471--489, 2010.

\bibitem[{Schneider} and {Lillo}(2016)]{schneider2016cross}
M.~{Schneider} and F.~{Lillo}.
\newblock {Cross-impact and no-dynamic-arbitrage}.
\newblock arXiv:1612.07742, 2016.

\bibitem[Sch{\"o}neborn(2016)]{schoneborn2016adaptive}
T.~Sch{\"o}neborn.
\newblock Adaptive basket liquidation.
\newblock \emph{Finance and Stochastics}, 20\penalty0 (2):\penalty0 455--493,
  2016.

\bibitem[T{\'o}th et~al.(2011)T{\'o}th, Lemp\'eri\`ere, Deremble,
  De~Lataillade, Kockelkoren, and Bouchaud]{toth2011anomalous}
B.~T{\'o}th, Y.~Lemp\'eri\`ere, C.~Deremble, J.~De~Lataillade, J.~Kockelkoren,
  and J.-P. Bouchaud.
\newblock Anomalous price impact and the critical nature of liquidity in
  financial markets.
\newblock \emph{Physical Review X}, 1\penalty0 (2):\penalty0 021006, 2011.

\bibitem[Wang(2017)]{wang2017trading}
S.~Wang.
\newblock Trading strategies for stock pairs regarding to the cross-impact
  cost.
\newblock arXiv:1701.03098, 2017.

\bibitem[Wang and Guhr(2016)]{wang2016microscopic}
S.~Wang and T.~Guhr.
\newblock Microscopic understanding of cross-responses between stocks: a
  two-component price impact model.
\newblock Available at \url{https://ssrn.com/abstract=2892266}, 2016.

\bibitem[Wang et~al.(2015)Wang, Sch{\"a}fer, and Guhr]{wang2015price}
S.~Wang, R.~Sch{\"a}fer, and T.~Guhr.
\newblock Price response in correlated financial markets: empirical results.
\newblock arXiv:1510.03205, 2015.

\end{thebibliography}

\end{document}